\documentclass[twocolumn,trackchanges]{aastex61}
\usepackage{listings}
\received{2017 August 3}
\revised{2017 October 15}
\accepted{2017 October 23}
\shorttitle{An Upper Bound on the Masses of Planets}
\shortauthors{Schlaufman}

\begin{document}

\title{Evidence of an Upper Bound on the Masses of Planets and its
Implications for Giant Planet Formation}

\correspondingauthor{Kevin C. Schlaufman}
\email{kschlaufman@jhu.edu}

\author[0000-0001-5761-6779]{Kevin C. Schlaufman}
\affiliation{Department of Physics and Astronomy \\
Johns Hopkins University \\
3400 North Charles Street \\
Baltimore, MD 21218, USA}

\begin{abstract}

\noindent
Celestial bodies with a mass of $M \approx 10~M_{\mathrm{Jup}}$
have been found orbiting nearby stars.  It is unknown whether these
objects formed like gas-giant planets through core accretion or like
stars through gravitational instability.  I show that objects with $M
\lesssim 4~M_{\mathrm{Jup}}$ orbit metal-rich solar-type dwarf stars,
a property associated with core accretion.  Objects with $M \gtrsim
10~M_{\mathrm{Jup}}$ do not share this property.  This transition
is coincident with a minimum in the occurrence rate of such objects,
suggesting that the maximum mass of a celestial body formed through
core accretion like a planet is less than $10~M_{\mathrm{Jup}}$.
Consequently, objects with $M \gtrsim 10~M_{\mathrm{Jup}}$ orbiting
solar-type dwarf stars likely formed through gravitational instability
and should not be thought of as planets.  Theoretical models of giant
planet formation in scaled minimum-mass solar nebula Shakura--Sunyaev
disks with standard parameters tuned to produce giant planets predict
a maximum mass nearly an order of magnitude larger.  To prevent newly
formed giant planets from growing larger than $10~M_{\mathrm{Jup}}$,
protoplanetary disks must therefore be significantly less viscous or of
lower mass than typically assumed during the runaway gas accretion stage
of giant planet formation.  Either effect would act to slow the Type I/II
migration of planetary embryos/giant planets and promote their survival.
These inferences are insensitive to the host star mass, planet formation
location, or characteristic disk dissipation time.

\end{abstract}


\section{Introduction}

Celestial bodies with a mass of $M \approx 10~M_{\mathrm{Jup}}$
have been found orbiting nearby stars as well as floating freely in
star-forming regions and the field \citep[e.g.,][]{lat89,zap00,kir06}.
Since it is currently impossible to determine the origin of a given
$10~M_{\mathrm{Jup}}$ object, it is unknown whether these bodies
formed like gas-giant planets through core accretion or like stars
through gravitational instability.  If it were practical to infer the
typical mass separating objects formed through core accretion from
those formed through gravitational instability, it would be possible
to derive otherwise unobtainable constraints on planet formation
models as well as the structure and properties of planet-forming disks
\citep[e.g.,][]{tan07,tan16}.

The existence of $10~M_{\mathrm{Jup}}$ objects both in orbit around
stars and in isolation, combined with the ambiguity of their formation,
has made it difficult to define a clear mass upper limit for planets.
The IAU Working Group on Extrasolar Planets adopted the working definition
of a planet as an object with a true mass below the limiting mass for
thermonuclear fusion of deuterium that orbits a star or stellar remnant.
On the other hand, substellar objects with true masses above the limiting
mass for thermonuclear fusion of deuterium were defined as brown dwarfs,
regardless of how they formed or where they are located \citep{bos07}.
At solar composition, the minimum mass to significantly burn deuterium
is thought to be near $13~M_{\mathrm{Jup}}$ independent of the formation
channel \citep[e.g.,][]{gro73,sau96,bur97,spi11,bod13}.  Other definitions
for giant planets referencing their intrinsic properties have also been
proposed \citep[e.g.,][]{hat15}.

The definition of planets as objects with $M \lesssim 13~M_{\mathrm{Jup}}$
is problematic for several reasons.  The critical mass for deuterium
burning is not a step function, so it is also necessary to arbitrarily
specify the fraction of the initial deuterium burned to uniquely define
the critical mass.  Since the critical mass depends on the composition,
at $M \approx 13~M_{\mathrm{Jup}}$ a metal-poor object would be a planet
while a metal-rich object would be a brown dwarf.  Because the mass
required for deuterium fusion depends on the internal properties of an
object, its calculation is necessarily dependent on imperfect models.
Finally, the definition does not take into account the potentially unique
ways such objects are formed.

The elegance of a formation-based planet definition has been broadly
recognized \citep[e.g.,][]{sch11,wri11,chab14}.  Under a formation-based
definition, planets are celestial bodies that form through core
accretion \citep{pol96,hub05}.  Conversely, brown dwarfs and stars form
through direct gravitational collapse, either at the disk or core scale
\citep[e.g.,][]{ada89,bat03,bat12,kra16}.  Despite the appeal of this
definition, it has not been put into practice because of the difficulty
in observationally determining the origin of individual celestial bodies
orbiting distant stars.

Celestial bodies that form through core accretion and gravitational
instability can be separated statistically though.  Giant planets
with a mass of $M \sim 1~M_{\mathrm{Jup}}$ preferentially orbit dwarf
stars that are metal rich.  This fact is thought to be indicative of
formation through core accretion \citep{san04,fis05,soz06,soz09}.
This observational inference is supported by theoretical exoplanet
population synthesis calculations, which suggest that the most massive
objects formed through core accretion should be exclusively found around
the most metal-rich stars \citep{mor12}.  In contrast, gravitational
instability is thought to occur with equal efficiency regardless of
gas-phase metallicity \citep{bos02,bat14}.  In accord with this idea,
the occurrence of low-mass stars with $M \sim 100~M_{\mathrm{Jup}}$
orbiting more massive stars like planets is independent of metal abundance
\citep{lat02,car03}.  The maximum mass at which celestial bodies no longer
preferentially orbit metal-rich solar-type dwarf stars can therefore
be used to separate massive planets from brown dwarfs and establish the
mass of the largest objects that can be formed through core accretion.

It has been impossible to confidently make this statistical separation
in the past due to the lack of giant planets and brown dwarfs
with mass estimates unaffected by the Doppler $\sin{i}$ degeneracy
orbiting solar-type dwarf stars with homogeneous stellar parameters.
That comparison is now possible.  In this paper, I calculate the mass
at which transiting objects no longer preferentially orbit metal-rich
solar-type dwarf stars.  The transition occurs somewhere in the range
of $4~M_{\mathrm{Jup}} \lesssim M \lesssim 10~M_{\mathrm{Jup}}$,
suggesting that objects with $M \lesssim 10~M_{\mathrm{Jup}}$ form
through core accretion like giant planets, while objects with $M \gtrsim
10~M_{\mathrm{Jup}}$ form like stars through gravitational instability.
I describe my sample definition in Section 2.  I detail my analysis
procedures and compare my results to a model of the runaway gas accretion
stage of giant planet formation in Section 3.  I discuss the overall
results and implications in Section 4 and conclude by summarizing my
findings in Section 5.

\section{Sample Definition}

I would like to calculate the mass at which secondary companions (e.g.,
planets, brown dwarfs, and low-mass stars) no longer preferentially orbit
metal-rich solar-type dwarf primaries.  I target systems where the minimum
mass, $M_2 \sin{i}$, of the secondary has been inferred with the Doppler
technique and the inclination, $i$, of the orbit is known to be close
to $90^{\circ}$ because of the observed transit.  That has two major
advantages.  First, the minimum mass inferred for the secondary in each
system from the Doppler measurement is very close to the true secondary
mass because the observed transit ensures that $i \approx 90^{\circ}
\Rightarrow \sin{i} \approx 1$.  Second, the observation of both the
Doppler and transit signals virtually guarantees that the secondary is
real and not a false positive.

Using the application programming interface (API) call in Appendix A,
I select from the NASA Exoplanet Archive all of the confirmed low-mass
secondaries that have $M_2 \geq 0.1~M_{\mathrm{Jup}}$ and have been
detected with both the Doppler and transit techniques.  I cross match
that sample on R.A. and decl. with the catalog of homogeneously
derived exoplanet host star stellar parameters from SWEET-Cat
\citep{san13,and17}.  Though I did not explicitly restrict the sample
to solar-type dwarf primaries, the requirements imposed above produce a
sample of primary stars with $4500~\mathrm{K} \lesssim T_{\mathrm{eff}}
\lesssim 7000~\mathrm{K}$ and $\log{g} \gtrsim 4.0$.  I rescale all of
the secondary masses from the NASA Exoplanet Archive sample using the
uniformly calculated SWEET-Cat primary star mass scale.  I include in
Table~\ref{tbl-1} the secondary masses and primary metallicities of the
resulting sample of 119 systems.  Because of the transit requirement,
95\% of the objects in Table~\ref{fig01} have an orbital period of $P <
10$ days and therefore a semimajor axis of $a \lesssim 0.1$ AU.

I compile from the literature a sample of brown dwarf and low-mass
star secondaries with $M_2 \lesssim 300~M_{\mathrm{Jup}} \approx
0.30~M_{\odot}$ that have Doppler-inferred masses and that transit
solar-type dwarf primaries.  I include in Table~\ref{tbl-2} the
secondary masses and primary metallicities of the resulting sample of
27 systems.  Because of the transit requirement, 95\% of the objects in
Table~\ref{fig02} have $P < 50$ days and therefore $a \lesssim 0.25$ AU.
This combined sample of 146 giant planets, brown dwarfs, and low-mass
stars that have Doppler-inferred masses and transit solar-type primaries
(most with homogeneously derived stellar parameters) is the best sample
available to calculate the mass at which low-mass secondary companions
no longer preferentially orbit metal-rich solar-type primaries.

Selection effects are unlikely to be present in this sample.  In a
signal-to-noise limited transit survey at a given period the number of
systems discovered, $N$ scales as $N \propto R_{1}^{-7/2} R_{2}^{6}$,
where $R_{1}$ is primary radius and $R_{2}$ is the secondary radius
\citep{pep03}.  All of the brown dwarfs and low-mass stars as well as
almost all of the giant planets in this sample were discovered with
the transit technique.  They all orbit solar-type host stars with
similar radii, so $R_{1} \approx 1~R_{\odot}$ for all 146 systems in
the sample considered here.  Because giant planets and brown dwarfs
are both significantly supported by degeneracy pressure, radius is a
weak function of mass in the interval of $1~M_{\mathrm{Jup}} \lesssim M
\lesssim 80~M_{\mathrm{Jup}}$ \citep[e.g.,][]{zap69,bur93}.  Consequently,
all 146 systems in this sample have $R_{2} \approx 1~R_{\mathrm{Jup}}$.
The Doppler technique can easily detect the orbital motion of the primary
in all 146 systems in this sample, so it can be used to estimate the
mass of the secondary in each system.  For these reasons, transit surveys
are equally complete for objects across the mass range of this sample.

Even though the period distribution of the transiting secondaries in
this sample is biased toward short periods, that bias will not affect
my goal of calculating the mass at which secondary companions no longer
preferentially orbit metal-rich solar-type dwarf primaries.  Both the
preference of giant planets for metal-rich primaries and the indifference
of low-mass stars to primary metallicity are independent of the period:
\citet{fis05} showed that giant planets preferentially orbit metal-rich
stars at all periods while \citet{lat02} demonstrated that the occurrence
and period distributions of metal-poor and metal-rich binary systems
are indistinguishable.  Accordingly, there is no reason to believe that
a similar analysis on a sample of the more common intermediate-period
secondaries would produce a qualitatively different conclusion.

\section{Analysis}

With the goal of identifying the mass at which low-mass secondaries no
longer preferentially orbit solar-type dwarf primaries, I analyze the
146 systems in Tables~\ref{tbl-1} and~\ref{tbl-2} in two independent
ways.  First, I apply clustering algorithms to the data in the
$\log_{10}$(secondary mass)--primary metallicity plane.  Since each
secondary formed either through core accretion or gravitational
instability, I specify in advance that each algorithm should put each
secondary in one of two clusters.  The mass that forms the border
between the two clusters is the mass at which secondaries no longer
preferentially occur around metal-rich solar-type dwarf stars, which I
interpret as the maximum mass of celestial bodies that form like planets
through core accretion.  Second, I take advantage of the theoretical
prediction of \citet{mor12} that massive objects formed by core accretion
should only occur around the most metal-rich primaries.  To that end,
I calculate the median metallicity of the sample as a function of
secondary mass in a kernel of constant width.  I interpret the point
at which the moving median metallicity drops below the smallest value
seen in the planet mass range as the maximum mass of objects that form
like planets, as massive objects formed by core accretion should orbit
stars more metal-rich than that of the typical giant planet--host star.
I detail these analyses in the following two subsections.  I then use a
semi-analytic model of the runaway gas accretion stage of giant planet
formation to interpret the results in the last subsection.

\subsection{Clustering Analysis}

\begin{figure*}[ht!]
\plotone{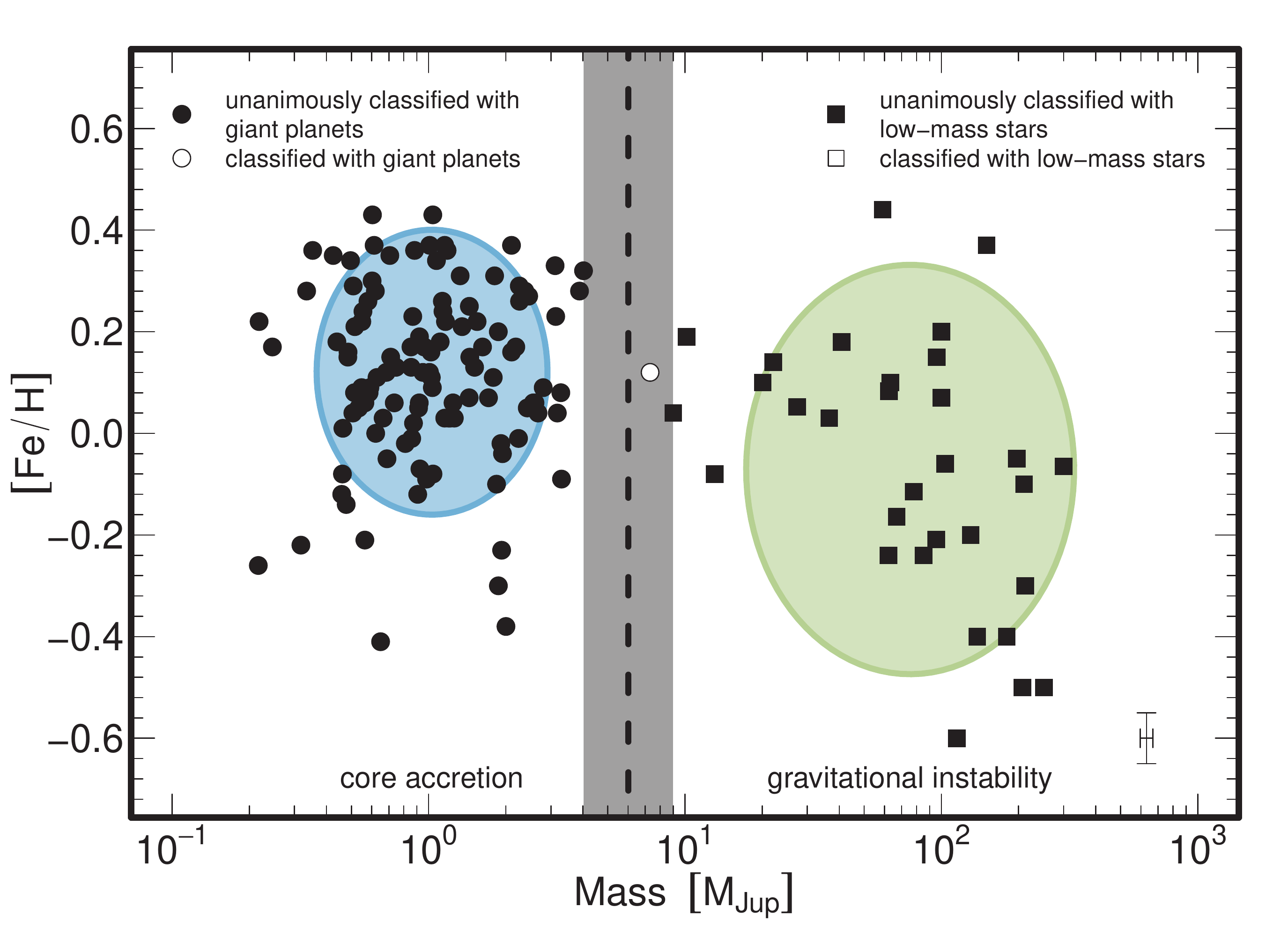}
\caption{Secondary mass--primary metallicity distribution for giant
planets, brown dwarfs, and low-mass stars that transit solar-type dwarf
stars.  Objects classified unanimously by all three of the clustering
paradigms are plotted as solid symbols, with circles for planets and
squares for low-mass stars.  Objects with classes agreed upon by two of
the three clustering paradigms are plotted as open symbols, with circles
for planets and squares for low-mass stars.  The gray shading separates
the unanimous classification regions.  The black dashed line is the
midpoint of the gray region.  The blue and green regions are the best-fit
two-component Gaussian mixture model, while the black cross in the lower
right corner of the plot indicates the typical uncertainty.\label{fig01}}
\end{figure*}

I use clustering algorithms from each of three classical clustering
paradigms that span the space of modern clustering approaches:
connectivity-based hierarchical clustering \citep{mur85}, centroid-based
$k$-means clustering \citep{har79}, and distribution-based Gaussian-model
clustering \citep{fra02,fra12,sch15}.  I summarize the key aspects of
each algorithm below and provide the details in Appendix B.

The hierarchical clustering algorithm seeks to identify the maximally
connected subgraph of a set of objects.  It starts with a matrix of
dissimilarity between all objects in a set.  It then identifies the
pair of objects with the smallest dissimilarity, combines them, and
updates the dissimilarity of all remaining objects in the set with the
minimum dissimilarity to either of the combined objects.  It continues
this process until only two objects remain.  This algorithm requires
the specification of a dissimilarity metric, and I use the Euclidean
distance between each object.

The $k$-means clustering algorithm seeks to minimize the total
dissimilarity between the ensemble of objects in each cluster and the
cluster center.  It starts by randomly placing a set of cluster centers
in the space defined by the objects to be clustered.  It then identifies
the closest cluster center to each object and assigns that object to that
cluster.  The mean value of the ensemble of objects in each cluster is
then set to be that cluster's center on the next iteration.  The process
continues until the movement of cluster centers on each update step
is small.  This algorithm requires the specification of a dissimilarity
metric, and I use the Euclidean distance between each object.

The Gaussian-model clustering algorithm seeks to identify the mixture
of Gaussians that has the highest likelihood of having generated the
objects to be clustered.  It starts by randomly choosing the mean and
variance of each Gaussian as well as the contribution of each Gaussian
to the mixture.  It then calculates the probability that each observation
was produced by one component of the mixture.  The object is assigned to
the component most likely to have generated it and the mean and variance
of each component are estimated based on the objects assigned to it.
The proportion of each component is the fraction of objects assigned
to it.  This process is repeated until the change in model parameters
is small.

To account for the observational uncertainty in each secondary mass
and primary metallicity, I use a Monte Carlo simulation.  First, I
generate $10^{3}+1$ realizations of each secondary mass and primary
metallicity from the observational uncertainties on those quantities
and then apply all three clustering algorithms to each realization.
For each algorithm, I aggregate the classifications across all
iterations and assign to each system the most frequently determined
classification.  The final classification of each system is then the
majority classification among the three different algorithms.  I plot
the result of this calculation in Figure~\ref{fig01}.  All systems with a
secondary mass of $M_2 < 4~M_{\mathrm{Jup}}$ are unanimously classified
with giant planets, while all systems with a secondary mass of $M_2 >
10~M_{\mathrm{Jup}}$ are unanimously classified with low-mass stars.
A two-sample Kolmogorov--Smirnov test on the overall sample split at $M_2
= 10~M_{\mathrm{Jup}}$ shows that the chance that the two metallicity
distributions were produced by the same parent distribution is only $p
= 4.45 \times 10^{-3}$, or about $3.3\sigma$.\footnote{The $p$-values
produced by the Kolmogorov--Smirnov test are not generated by comparison
with a Gaussian distribution, so I only give a numerical value of $\sigma$
to aid in the comparison of this result to those of previous studies.}

\subsection{Moving Median Analysis}

The clustering analysis described above implicitly makes use of
the relative occurrence rates of giant planets, brown dwarfs, and
low-mass stars.  An alternative approach that does not implicitly use the
occurrence rate is to calculate the median metallicity as a function of
secondary mass in a kernel of a constant width.  Since celestial bodies
with $M \gtrsim 10~M_{\mathrm{Jup}}$ formed via core accretion are
thought to occur only around the most metal-rich stars \citep{mor12},
the mass at which the moving median metallicity moves below the lowest
moving median metallicity observed in the giant planet region should also
correspond to the maximum mass of the objects formed by core accretion.

\begin{figure}
\plotone{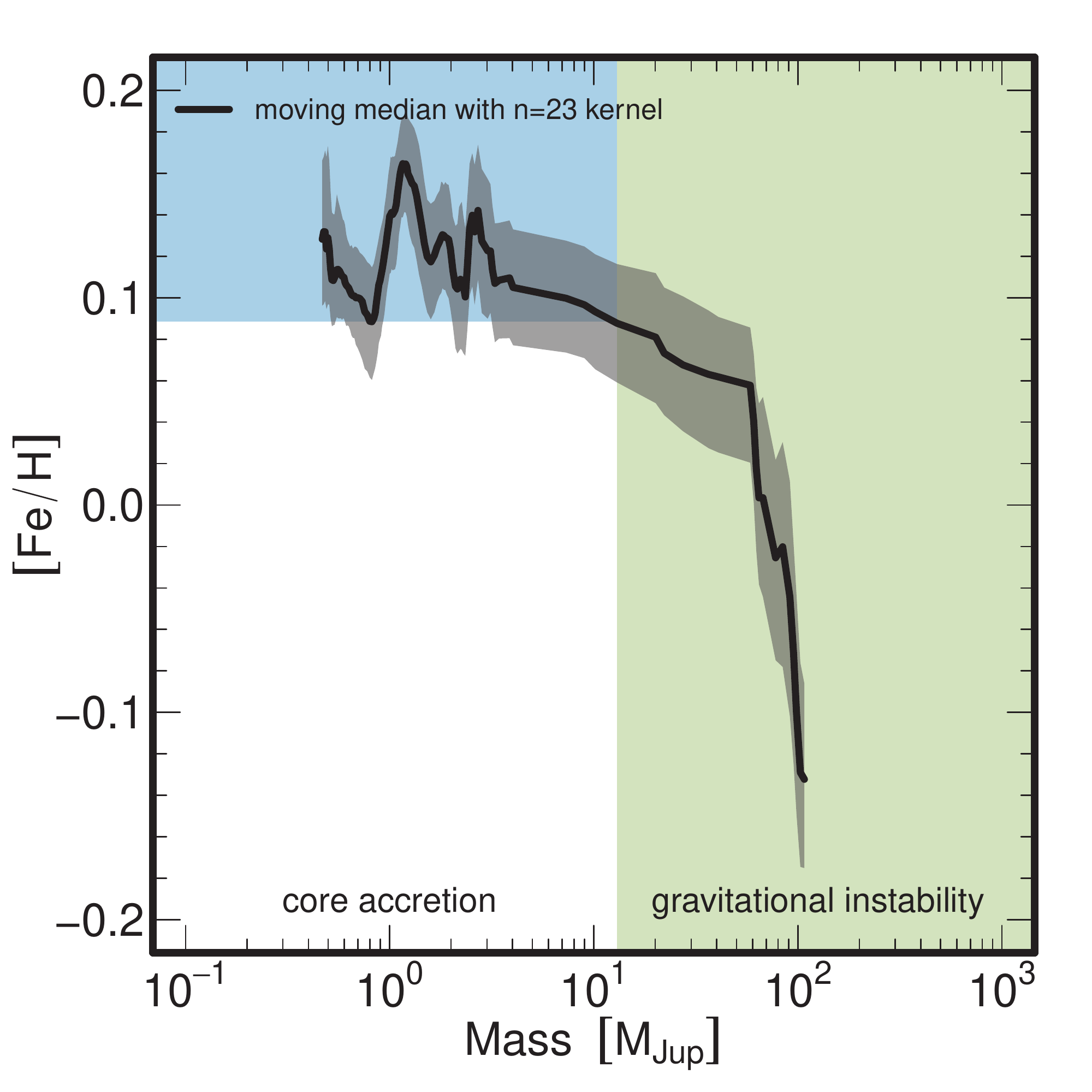}
\caption{Moving median metallicity as a function of secondary mass.
The black line is the median metallicity in a kernel of width $n = 23$
centered on the given mass.  The gray region is the 1$\sigma$ uncertainty
in the moving median.  The blue rectangle indicates the range of median
metallicity observed in the giant planet region.  The green rectangle
shows the range in mass where the moving median is below that seen
in the giant planet region (i.e., $M \lesssim 4~M_{\mathrm{Jup}})$.
The boundary is near $10~M_{\mathrm{Jup}}$.\label{fig02}}
\end{figure}

To account for the observational uncertainty in each secondary mass
and primary metallicity, I use a Monte Carlo simulation.  First, I
generate $10^{3}+1$ realizations of each secondary mass and primary
metallicity from the observational uncertainties on those quantities
and then calculate the moving median metallicity of each sample as a
function of mass.  I use a kernel of width $n = 23$ objects, as this is
the highest mass resolution that permits the measurement of the median
metallicity with a similar precision as for the individual metallicity
estimates ($\sigma_{\mathrm{[Fe/H]}} \approx 0.05$ dex).  I plot this
result in Figure~\ref{fig02}.  Again, the point at which secondaries no
longer preferentially orbit metal-rich solar-type dwarf-star primaries
occurs at $M_2 \approx 10~M_{\mathrm{Jup}}$.  This result is similar for
all kernel widths that permit the measurement of the median metallicity
with a similar precision as for the individual metallicity estimates.

\subsection{Parameter Study of the Last Stage of Giant Planet Formation}

The results in Sections 3.1 and 3.2 show that low-mass secondaries no
longer preferentially orbit metal-rich solar-type dwarf stars when $M_2
\gtrsim 10~M_{\mathrm{Jup}}$.  Those inferences support the idea that
the most massive objects formed around solar-type stars like planets
through core accretion have $M \approx 10~M_{\mathrm{Jup}}$.  I now use
that inference to quantify the range of protoplanetary disk properties
that could be expected to reproduce this result using the semi-analytic
model of \citet{tan16}.

\citet{tan16} modeled the last stage of giant planet formation (runaway
envelope accretion) in a globally evolving protoplanetary disk using
the \citet{sha73} model for disk viscosity and a self-similar solution
for the disk's global evolution \citep[e.g.,][]{lyn74,har98}.  The key
parameters of the model are as follows:

\begin{enumerate}
\item
The location of planet formation, $a_p$.
\item
The disk viscosity, $\nu = \alpha c h$, where $\alpha$ is a unitless
parameter independent of radius and time, $c$ is the sound speed of disk
gas, and $h$ is the disk scale height.
\item
The initial mass of the disk, $M_{\mathrm{disk}}$, is assumed to scale
linearly with a host star mass, $M_{\ast}$ and overall normalization,
$f_{\Sigma}$.  Here $f_{\Sigma} = 1$ corresponds to the minimum-mass
solar nebula of \citet{hay85}.
\item
The characteristic exponential disk depletion time, $\tau_{\mathrm{dep}}$.
\end{enumerate}

\noindent
The model is informed by the latest numerical results showing that
gaps cleared by giant planets are much shallower than previously
thought \citep[e.g.,][]{duf13,fun14}.  The model does not account for
possible photoevaporation of disk gas due to far-ultraviolet radiation
\citep[e.g.,][]{gor09}, as the rate of mass loss expected is still
uncertain by more than an order of magnitude \citep[e.g.,][]{ale14}.

\begin{figure*}
\plotone{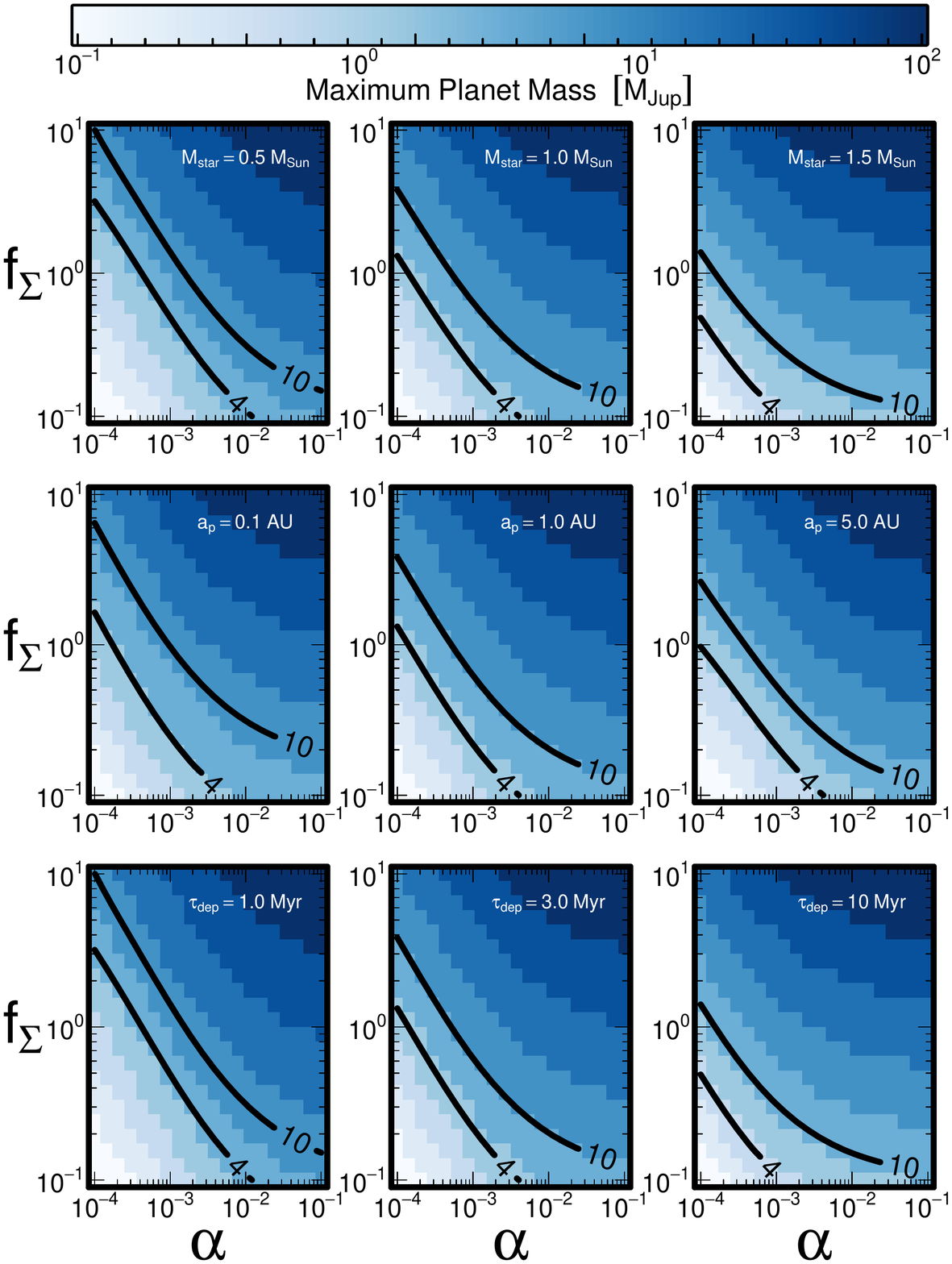}
\caption{Maximum mass attainable by a giant planet formed by core
accretion according the model of \citet{tan16} as a function of the
viscosity parameter $\alpha$ of \citet{sha73} and minimum-mass solar
nebular scaling factor, $f_{\Sigma}$.  Dark colors are indicative of
high-mass planets, while lighter colors are indicative of low-mass
planets.  The labeled contours delimit the region of maximum mass
permitted by the analyses presented in Sections 3.1 and 3.2.  The top row
varies the host star mass, $M_{\ast}$, the middle row varies the location
of planet formation, $a_p$, and the bottom row varies the characteristic
disk depletion time, $\tau_{\mathrm{dep}}$.\label{fig03}}
\end{figure*}

I have implemented the \citet{tan16} model and used it to conduct a
parameter study of the relationship between the maximum attainable mass
of a giant planet and the model parameters $a_p$, $\alpha$, $M_{\ast}$,
$f_{\Sigma}$, and $\tau_{\mathrm{dep}}$.  If not otherwise indicated, I
assume that $a_p$ = 1 au, $M_{\ast} = 1~M_{\odot}$, $\tau_{\mathrm{dep}}
= 3 \times 10^6$ yr, the initial disk radius is $R_0 = 200$ AU, and the
seed embryo mass is $M_{\mathrm{embryo}} = 10~M_{\oplus}$.  I plot the
result in Figure~\ref{fig03}.  According to the model, a maximum giant
planet mass in the range of $4~M_{\mathrm{Jup}} \lesssim M \lesssim
10~M_{\mathrm{Jup}}$ can be accommodated only in a narrow corridor of the
$\alpha$--$f_{\Sigma}$ plane from $\alpha \sim 10^{-4}$ and $f_{\Sigma}
\sim 10^{0.5}$ to $\alpha \sim 10^{-2}$ and $f_{\Sigma} \sim 10^{-1}$.
This result is only weakly dependent on the mass of the primary star, the
location of planet formation, or the disk depletion time.  These results
are unique in that they indirectly constrain the viscosities and masses of
long since dissipated disks during their epoch of giant planet formation.

\section{Discussion}

I have shown that celestial bodies with $M \lesssim 10~M_{\mathrm{Jup}}$
orbit metal-rich solar-type dwarf stars, a property thought to indicate
formation through core accretion \citep[e.g.,][]{san04,fis05}.  On the
other hand, celestial bodies with $M \gtrsim 10~M_{\mathrm{Jup}}$ do not
preferentially orbit metal-rich stars, so there is no reason to believe
that they form through core accretion.  Since gravitational instability
is thought to be independent of metallicity \citep[e.g.,][]{bos02},
it can easily accommodate this observation.

\begin{figure}
\plotone{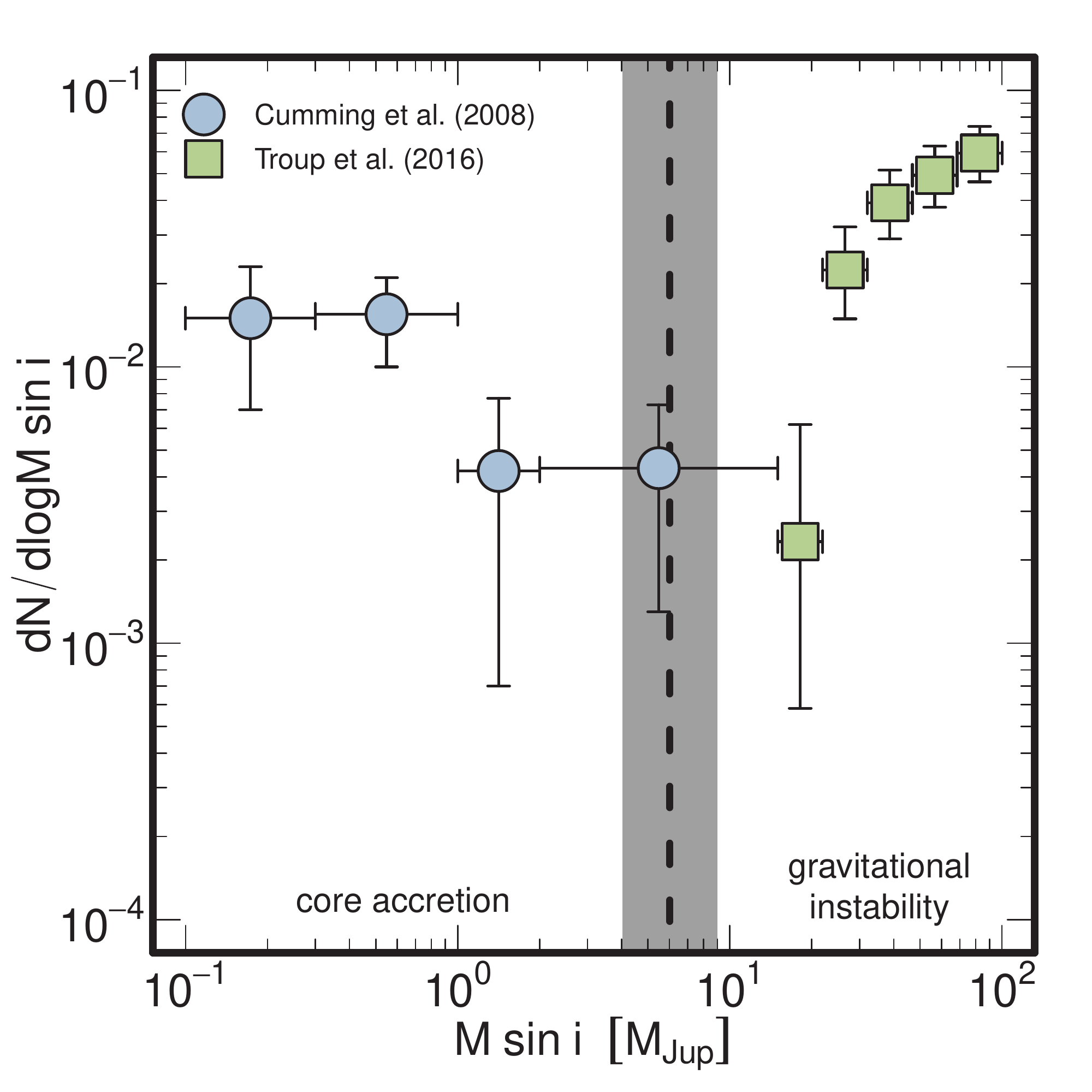}
\caption{Secondary occurrence rate as a function of secondary
minimum mass at $P < 100$ days.  The secondary occurrence rate in
the range of $0.1~M_{\mathrm{Jup}} < M\sin{i} < 15~M_{\mathrm{Jup}}$
is from \citet{cum08}, while the occurrence rate in the range of
$15~M_{\mathrm{Jup}} < M\sin{i} < 100~M_{\mathrm{Jup}}$ is from
\citet{tro16}.  The gray shading and black dashed line are the same
as in Figure~\ref{fig01}.  They separate the regions identified by the
clustering analysis.\label{fig04}}
\end{figure}

This boundary at $M \approx 10~M_{\mathrm{Jup}}$ between objects formed
via core accretion and gravitational instability is coincident with a
minimum in the occurrence rate of companions to solar-type dwarf stars
with $P < 100$ days as a function of minimum mass.  I plot the occurrence
rate of giant planets, brown dwarfs, and low-mass stars discovered with
the Doppler technique in Figure~\ref{fig04}.  The observation that the
upper limit on the mass of giant planets identified by the clustering
analysis occurs at the minimum in the occurrence rate of celestial bodies
orbiting solar-type dwarf stars supports the idea that core accretion
does not generally form objects with $M \gtrsim 10~M_{\mathrm{Jup}}$.

This boundary at $M \approx 10~M_{\mathrm{Jup}}$ between objects formed
via core accretion and gravitational instability is also supported by
interior models for three massive objects that straddle the boundary:
HAT-P-20 b \citep[$M \approx 7.3~M_{\mathrm{Jup}}$ ---][]{bak11}, HAT-P-2
b \citep[$M \approx 9.0~M_{\mathrm{Jup}}$ ---][]{bak07a}, and Kepler-75 b
\citep[$M \approx 10.1~M_{\mathrm{Jup}}$ ---][]{bon15}.  Interior models
indicate HAT-P-20 b and HAT-P-2 b have large amounts of heavy elements
as expected if they formed through core accretion, while Kepler-75 b
is consistent with the composition of its host star as expected if it
formed through gravitational instability \citep{lec09,tho16}.

\subsection{Implications for Planet Formation}

The results presented in Section 3 indicate that a $M \approx
10~M_{\mathrm{Jup}}$ maximum mass for objects formed by core accretion
can only be accommodated in a narrow band in the \citet{sha73} viscosity
parameter $\alpha$--minimum-mass solar nebular scaling factor $f_{\Sigma}$
plane from $\alpha \sim 10^{-4}$ and $f_{\Sigma} \sim 10^{0.5}$ to
$\alpha \sim 10^{-2}$ and $f_{\Sigma} \sim 10^{-1}$.  Observations
of T Tauri stars have been used to suggest that protoplanetary
disks can be well reproduced by models with $\alpha \sim 10^{-2}$
\citep[e.g.,][]{har98}.  At the same time, classical models of giant
planet formation though core accretion almost always invoke disks several
times more massive than that of the minimum-mass solar nebula---or
$f_{\Sigma} \gtrsim 3$---to ensure core formation before disk dissipation
\citep[e.g.,][]{pol96,iko00,ali05,hub05,liss09,mov10,dan14}.  In a disk
with $\alpha = 10^{-2}$ and $f_{\Sigma} = 3$, the \citet{tan16} model
predicts a giant planet maximum mass of $M \approx 60~M_{\mathrm{Jup}}$.
According to these results, either but not both, of these traditional
assumptions could be made during the runaway gas accretion stage of
giant planet formation.

If the $M \approx 10~M_{\mathrm{Jup}}$ threshold is the result of
low-viscosity or low-mass disks, then it would provide solutions
to the long-standing Type I and Type II migration problems.  Type I
migration occurs when planetary embryos with $M_\mathrm{embryo} \sim
1~M_{\oplus}$ that are not massive enough to open a gap in their
parent protoplanetary disks lose angular momentum and move inward
due to net negative torques from the disk.  Though the processes
responsible are complex and no model is yet completely satisfactory,
the simplest models suggest that the timescale for migration is very fast
\citep[e.g.,][]{war97,tan02}.  Exoplanet population synthesis calculations
assuming fast Type I migration are unable to reproduce the existence
of planets with $M \gtrsim 10~M_{\oplus}$ and $P \gtrsim 10$ days, so
Type I migration must be slower than the simplest calculations initially
suggested \citep[e.g.,][]{ida08,sch09}.  While many solutions have been
proposed,\footnote{See \citet{bar16} for a comprehensive review.} it has
been shown that Type I migration in disks with $\alpha \sim 10^{-4}$ is
sufficiently slow that planetary embryos will survive and potentially grow
into giant planets \citep[e.g.,][]{raf02,li09}.  Similarly, in disks with
$\alpha \sim 10^{-2}$, the Type I migration rate is thought to be linearly
dependent on disk surface density \citep[e.g.,][]{bar16}.  A minimum-mass
solar nebular scaling factor of $f_{\Sigma} \sim 10^{-1}$ would reduce
the Type I migration rate by an order of magnitude, exactly the amount
needed to reproduce the observed distribution of exoplanet properties
in population synthesis calculations \citep[e.g.,][]{ida08,sch09}.

Low-viscosity or low-mass disks would also solve the Type II migration
problem.  Type II migration occurs when planets that are massive
enough to open a gap in their parent protoplanetary disks are carried
into the close proximity of their host star on the disk viscous time
\citep{lin96}, possibly precluding the existence of long-period giant
planets.  In disks where the mass of the disk locally dominates the mass
of the planet, the Type II migration rate is inversely proportional to
the viscosity \citep[e.g.,][]{lin96,tan16}.  Similarly, in a low-mass
disk where the mass of the planet locally dominates the mass of the disk,
the Type II migration rate is inversely proportional to the disk mass
\citep[e.g.,][]{duf14}.  Consequently, low-viscosity or low-mass disks
promote the existence of long-period giant planets.

Though these results were derived strictly for solar-type dwarf stars
with short-period transiting companions, the analysis in Section 3.3
suggests that the inference about planet formation is not strongly
dependent on host-star mass, planet-formation location, or disk
dissipation timescale.  Indeed, similar inferences have been made using
the much-more distant solar system giant planets Jupiter and Saturn
by \citet{liss09} and \citet{tan16}.  It is also reassuring that this
upper limit of $10~M_{\mathrm{Jup}}$ on the mass of planets within 1
au of their host stars is consistent with the most advanced exoplanet
population synthesis calculations currently available \citep{ida13}.
Finally, support for low-viscosity or low-mass disks as the origin of
planetary systems has been found in many places from meteoritics in
the solar system to super-Earths and mini-Neptunes in the Kepler field
\citep[e.g.,][]{des17,lee17}.

\subsection{Comparison to Previous Studies of the Relationship between
Giant Planets, Brown Dwarfs, and Host-star Properties}

This study focused on giant planets, brown dwarfs, and low-mass stars
with Doppler-inferred masses transiting solar-type dwarf stars with
homogeneously estimated stellar parameters.  It showed that low-mass
secondaries with $M_2 < 10~M_{\mathrm{Jup}}$ preferentially orbit
metal-rich stars, while secondaries with $M_2 > 10~M_{\mathrm{Jup}}$
occur independent of host star metallicity.  The difference between
the metallicity distributions of objects in my sample with $M >
10~M_{\mathrm{Jup}}$ and $M < 10~M_{\mathrm{Jup}}$ is significant at
more than $3.3~\sigma$, and this result cannot be attributed solely to
the possible change in the occurrence rate of such bodies at $M \approx
4~M_{\mathrm{Jup}}$.  Since a preference for metal-rich host stars is
thought to indicate formation through core accretion, these results
suggest that the most massive objects that can be formed like planets
through core accretion have $M \approx 10~M_{\mathrm{Jup}}$.

Unlike all previous studies, these results are unaffected by the
Doppler $\sin{i}$ degeneracy, the presence of possible false positives
in the analysis sample, or the potentially weak or nonexistent
giant planet--host star metallicity correlation for giant stars
\citep[e.g.,][]{sad05,sch05,hek07,pas07,tak08,ghe10,mal13,mor13a,jof15,ref15}.
Because of its large sample and use of $M$ instead of $M\sin{i}$, it
has the best resolution in mass of all studies to date.  All of these
properties have enabled the identification of a difference between the
metallicity distributions of objects with $M > 10~M_{\mathrm{Jup}}$
and $M < 10~M_{\mathrm{Jup}}$ at a higher statistical significance than
any previous study of solar-type dwarfs.  It is also the first study to
couple its result to a detailed model of the runaway gas accretion stage
of giant planet formation to constrain the properties of protoplanetary
disks that are known to have formed giant planets.

The first studies seeking to compare giant planets and brown dwarfs
to explore their origin all focused on the orbital properties of brown
dwarfs.  \citet{gre06} examined the mass distribution of companions to
solar-type stars with $P < 5$ yr and determined that the minimum of the
distribution occurs at $M\sin{i} = 31^{+25}_{-18}~M_{\mathrm{Jup}}$.
Because there was no such minimum in the mass spectrum of free-floating
brown dwarfs, they speculated that migration and not a characteristic
mass scale of fragmentation in protoplanetary disks was responsible for
the apparent minimum.

\citet{sah11} used the combination of the Doppler and astrometric
techniques to remove the $\sin{i}$ degeneracy to confirm the
previous result, finding that low-mass brown dwarfs with $M \lesssim
25~M_{\mathrm{Jup}}$ cleanly separated from high-mass brown dwarfs
with $M \gtrsim 45~M_{\mathrm{Jup}}$.  In contrast to \citet{gre06},
they argued that the low-mass brown dwarfs may be massive planets
formed by core accretion.  They also showed that samples of brown dwarf
candidates discovered solely by the Doppler technique with $M\sin{i}
\gtrsim 45~M_{\mathrm{Jup}}$ are significantly contaminated by inclined
low-mass stars.\footnote{The frequent occurrence of low-mass stars in
Doppler-selected samples of high-mass brown dwarfs has also been seen
by \citet{wil16}.}

\citet{ma14} extended these results by demonstrating that brown dwarfs
with $M\sin{i} \lesssim 42.5~M_{\mathrm{Jup}}$ and $M\sin{i} \gtrsim
42.5~M_{\mathrm{Jup}}$ are unlikely to have the same eccentricity
distribution at the 2.1$\sigma$ level.  They suggested that the low-mass
brown dwarfs formed by disk fragmentation, while the high-mass brown
dwarfs formed by fragmentation at the core scale.  They also showed that
the giant planet and brown dwarf host stars were unlikely to have the same
metallicity distribution at the 3.3$\sigma$ level.  Selection effects in
the giant planet--host sample and inclined low-mass star contamination
in the brown dwarf sample likely affect these conclusions to some
degree though.

More recent efforts to investigate the differences between massive planets
and brown dwarfs have focused on host star metallicity and chemical
abundance.  Using a sample of solar-type dwarf stars unaffected by
selection effects or inclined low-mass star contamination, \citet{mat14}
found that the metallicities and $\alpha$-element abundances of brown
dwarf hosts are lower than the metallicities of giant planet--host stars
at the 1.7$\sigma$ and 1.8$\sigma$ levels.

\citet{mal17} presented a similar analysis to \citet{ma14} that used
homogeneously derived stellar parameters to show that low-mass brown
dwarf host are more metal-rich than high-mass brown dwarf hosts at the
2.4$\sigma$ level.  However, the difference was not significant if only
dwarf host stars were included or if a random distribution of inclination
was assumed.  The fact that the significance of their result is reduced
when accounting for a random distribution of inclination suggests that
low-mass stars on low-inclination orbits contaminated their sample.
In addition, their compete sample of brown dwarf hosts was significantly
more metal poor than their sample of giant planet--host stars.  At the
same time, their low-mass brown dwarfs were consistent with having the
same metallicity distribution as the giant planet--host stars, though
their $\alpha$-element distributions differed at the 2.4$\sigma$ level.

\citet{san17} studied the minimum-mass and metallicity distributions of
mostly Doppler-discovered objects with $M\sin{i} < 15~M_{\mathrm{Jup}}$,
$10~\mathrm{days} < P < 1825~\mathrm{days}$, and with homogeneous stellar
parameters available in the SWEET-Cat database.  Using both multivariate
Gaussian modeling and Kolmogorov--Smirnov tests, they argued that
planets below and above $M\sin{i} = 4~M_{\mathrm{Jup}}$ have distinct
metallicity distributions and therefore distinct formation channels.  In
particular, they suggested that the giant planets with $M\sin{i} \lesssim
4~M_{\mathrm{Jup}}$ formed via core accretion while the giant planets
with $M\sin{i} \gtrsim 4~M_{\mathrm{Jup}}$ formed via gravitational
instability.  Their multivariate Gaussian inference is mostly due to the
apparent cutoff in the giant planet occurrence distribution at $M\sin{i}
\approx 4~M_{\mathrm{Jup}}$.  In their Kolmogorov--Smirnov tests, the
distributions were distinct at the 3.3$\sigma$ significance substantially
because of the inclusion of giant stars with $M \gtrsim 1.5~M_{\odot}$.
The fact that restricting the sample to solar-type dwarf stars reduced
the significance of their inference to 1.6 $\sigma$ indicates that either
false positives in the giant star sample or the possibly diminished giant
planet occurrence--host star metallicity effect for giant stars may play
a role in the significant difference they identified in their sample
including both dwarf and giant host stars.  A critical mass of $M\sin{i} =
4~M_{\mathrm{Jup}}$ is also inconsistent with the heavy element enrichment
observed in HAT-P-20 b and HAT-P-2 b.  Nevertheless, my study confirms
the hint identified in \citet{san17} and presents the strongest evidence
to date that the maximum mass of objects formed via core accretion is in
the range of $4~M_{\mathrm{Jup}} \lesssim M \lesssim 10~M_{\mathrm{Jup}}$.

\subsection{A Formation-based Definition for Planets}

This analysis has shown that objects with $M \lesssim 10~M_{\mathrm{Jup}}$
preferentially orbit metal-rich solar-type dwarf stars.  That property
has been suggested to be a natural outcome of the core accretion model
of giant planet formation.  On the other hand, objects with $M \gtrsim
10~M_{\mathrm{Jup}}$ orbit stars spanning the whole range of the thin
disk metallicity distribution.  That property is shared with low-mass
stars, which must have formed by some sort of gravitational instability.

I propose that planets be defined as objects that orbit stars or stellar
remnants and have a true mass below the threshold at which low-mass
companions no longer preferentially orbit metal-rich solar-type dwarf
stars.  The analyses presented in Sections 3.1 and 3.2 suggest this
threshold is at $M \approx 10~M_{\mathrm{Jup}}$, but future data may
revise this estimate.  This definition has the same form as the working
definition adopted by the IAU Working Group on Extrasolar Planets, but
the maximum mass is now referenced to the maximum mass of objects formed
by core accretion.  This definition does not require the specification
of an arbitrarily amount of deuterium burning, and it does not depend
on an object's metallicity.  Additionally, it is independent of the
uncertain internal structure of objects at this mass scale.  Furthermore,
I propose that substellar objects with true masses above the threshold
be defined as brown dwarfs, regardless of where they are located.

\subsection{Predictions for Future Gaia Observations}

The results presented in Sections 3.1 and 3.2 lead to the prediction that
planets with $1~M_{\mathrm{Jup}} \lesssim M \lesssim 10~M_{\mathrm{Jup}}$
should preferentially be found around metal-rich solar-type dwarf
stars.  In contrast, brown dwarfs with $10~M_{\mathrm{Jup}} \lesssim M
\lesssim 80~M_{\mathrm{Jup}}$ should be found around solar-type dwarf
stars that span the metallicity range of the Milky Way's thin disk.
This prediction will soon be tested at an unprecedented scale.  By the
end of its planned five-year mission, the European Space Agency's
Gaia satellite is expected to discover astrometrically and characterize
$21,\!000\pm6000$ giant planets and brown dwarfs with $1~M_{\mathrm{Jup}}
< M < 15~M_{\mathrm{Jup}}$ and $1~\mathrm{AU} \lesssim a \lesssim
5~\mathrm{AU}$ \citep{per14}.  This sample of giant planet and brown
dwarf mass measurements will be nearly 100 times larger than the sample
currently available and extend to the more common intermediate-period
objects lacking in my sample.  For these reasons, Gaia will provide the
definitive test of this prediction that planets with $1~M_{\mathrm{Jup}}
\lesssim M \lesssim 10~M_{\mathrm{Jup}}$ should preferentially be
found around metal-rich solar-type dwarf stars, while brown dwarfs with
$10~M_{\mathrm{Jup}} \lesssim M \lesssim 80~M_{\mathrm{Jup}}$ should be
found around solar-type dwarf stars that span the metallicity range of
the Milky Way's thin disk.

\section{Conclusion}

Celestial bodies with $M \lesssim 10~M_{\mathrm{Jup}}$ preferentially
orbit metal-rich solar-type dwarf stars, while celestial bodies with
$M \gtrsim 10~M_{\mathrm{Jup}}$ do not preferentially orbit metal-rich
solar-type dwarf stars.  A preference for metal-rich host stars is
thought to be a property of objects formed like giant planets through
core accretion, while objects formed like stars through gravitational
instability should not prefer metal-rich primaries.  As a result,
these data suggest that core accretion rarely forms giant planets
with $M \gtrsim 10~M_{\mathrm{Jup}}$ and objects more massive than
$M \approx 10~M_{\mathrm{Jup}}$ should not be thought of as planets.
Instead, objects with $M \gtrsim 10~M_{\mathrm{Jup}}$ formed like
stars through gravitational instability.  An upper limit of $M \approx
10~M_{\mathrm{Jup}}$ to the mass of planets can only be accommodated in
either low-viscosity or low-mass protoplanetary disks.  In either case,
both Type I and Type II migration are an order of magnitude slower than
traditionally assumed.  For that reason, these results may point toward
the solution of both the Type I and Type II migration problems.  Finally,
these observations put the definition of a planet as a secondary with
$M \lesssim 10~M_{\mathrm{Jup}}$ formed via core accretion on a solid
observational basis for the first time.

\acknowledgments
I thank Andy Casey, Greg Laughlin, Margaret Moerchen, Josh Simon, and
Josh Winn for insightful comments.  This research has made use of NASA's
Astrophysics Data System Bibliographic Services, the SIMBAD database,
operated at CDS, Strasbourg, France \citep{wen00}, and the NASA Exoplanet
Archive, which is operated by the California Institute of Technology,
under contract with the National Aeronautics and Space Administration
under the Exoplanet Exploration Program.

\vspace{5mm}
\software{\texttt{mclust} \citep{fra12},
          \texttt{R} \citep{r17},
          \texttt{TOPCAT} \citep{tay05}  
          }

\appendix

\section{API Call for Sample Selection}

The following URL can be used to reproduce
my initial selection with the API available from the
\href{https://exoplanetarchive.ipac.caltech.edu/docs/program\_interfaces.html}{NASA
Exoplanet Archive}:

\begin{lstlisting}
http://exoplanetarchive.ipac.caltech.edu/cgi-bin/nstedAPI/nph-nstedAPI?&table=exoplanets&select=pl_hostname,ra,dec,st_mass,st_masserr1,st_masserr2,st_metfe,st_metfeerr1,st_metfeerr2,pl_name,pl_letter,pl_bmassj,pl_bmassjerr1,pl_bmassjerr2,pl_massj,pl_massjerr1,pl_massjerr2,pl_disc_refname,pl_disc_reflink,pl_def_refname,pl_def_reflink&where=pl_tranflag=1 and pl_rvflag=1 and pl_massj>=0.1&order=pl_bmassj
\end{lstlisting}

\section{Details of Clustering Algorithms}

\subsection{Hierarchical clustering}

I use the \texttt{hclust} function in \texttt{R} for hierarchical
clustering \citep{r17}.  That implementation follows the algorithm of
\citet{mur85}:

\begin{enumerate}
\item
Calculate the set of $n(n-1)/2$ dissimilarities between $n$ objects.
\item
Identify the smallest dissimilarity $d_{ik}$.
\item
Combine objects $i$ and $k$.  That is, replace the two objects with a
new object, $i \bigcup k$, and update all other dissimilarities such
that for all objects $j \neq i,k$ the dissimilarity $d_{i \bigcup k,j}
= \min{(d_{i,j},d_{k,j})}$.  Delete the dissimilarities $d_{i,j}$
and $d_{k,j}$.
\item
Return to Step 2 while the number of objects remaining is greater
than two.
\end{enumerate}

\noindent
This algorithm requires the specification of a dissimilarity metric,
and I use the Euclidean distance between each object.

\subsection{$k$-means Clustering}

I use the \texttt{kmeans} function in \texttt{R} for $k$-means clustering
\citep{r17}.  That implementation follows the algorithm of \citet{har79}:

\begin{enumerate}
\item
Pick $k$ objects from the $n$ objects to be clustered.
\item
Assign each object to a cluster by placing each it in the cluster,
$k_{i}$, that has the centroid with the smallest dissimilarity.
\item
Recalculate the centroid of each cluster.
\item
Return to Step 2, unless changes in the cluster centroids are small.
\end{enumerate}

\noindent
This algorithm requires the specification of a dissimilarity metric,
and I use the Euclidean distance between each object.

\subsection{Gaussian-model Clustering}

I use the \texttt{mclust} package in \texttt{R} for Gaussian-model
clustering \citep{fra12,r17}.  The algorithm is considerably more complex
than either the hierarchical or $k$-means algorithms, so the interested
reader should see \cite{fra02} for the details of the algorithm.

%\nocite{*}

\begin{longrotatetable}
\begin{deluxetable*}{ccllDDcc}
\tabletypesize{\tiny}
\tablecaption{Giant Planets and Brown Dwarfs that have Doppler-inferred Masses and Transit Solar-type Stars with Homogeneous Stellar Parameters from SWEET-Cat\label{tbl-1}}
\tablehead{
\colhead{System} & \colhead{Simbad} & \colhead{R.A.} & \colhead{Decl.} & \twocolhead{Mass} & \twocolhead{Metallicity} & \colhead{Discovery} & \colhead{Stellar Parameter} \\
\colhead{} & \colhead{Name} & \colhead{} & \colhead{} & \twocolhead{} & \twocolhead{} & \colhead{Reference} & \colhead{Reference} \\
\colhead{} & \colhead{} & \colhead{(h m s)} & \colhead{(d m s)} & \twocolhead{($M_{\mathrm{Jup}}$)} & \twocolhead{} & \colhead{} & \colhead{}
}
\decimals
\startdata
CoRoT-8 & CoRoT-8 & 19 26 21.243 & +01 25 35.17 &  0.22^{+0.03}_{-0.03} &  0.22^{+0.11}_{-0.11} & \citet{bor10} & \citet{mor13b}\\
HAT-P-12 & GSC 03033-00706 & 13 57 33.48 & +43 29 36.7 &  0.22^{+0.01}_{-0.01} & -0.26^{+0.06}_{-0.06} & \citet{har09} & \citet{and17}\\
WASP-29 & TYC 8015-1020-1 & 23 51 31.085 & $-$39 54 24.26 &  0.25^{+0.02}_{-0.02} &  0.17^{+0.05}_{-0.05} & \citet{hel10} & \citet{mor13b}\\
WASP-21 & GSC 01715-00679 & 23 09 58.254 & +18 23 45.88 &  0.32^{+0.01}_{-0.01} & -0.22^{+0.04}_{-0.04} & \citet{bou10} & \citet{mor13b}\\
WASP-63 & TYC 7612-556-1 & 06 17 20.7489 & $-$38 19 23.773 &  0.33^{+0.03}_{-0.03} &  0.28^{+0.05}_{-0.05} & \citet{hel12} & \citet{mor13b}\\
HD 149026 & HD 149026 & 16 30 29.6185 & +38 20 50.308 &  0.35^{+0.01}_{-0.01} &  0.36^{+0.05}_{-0.05} & \citet{sat05} & \citet{amm09}\\
WASP-94 A & WASP-94A & 20 55 07.946 & $-$34 08 08.00 &  0.42^{+0.03}_{-0.03} &  0.35^{+0.03}_{-0.03} & \citet{nev14} & \citet{and17}\\
WASP-67 & TYC 6307-1388-1 & 19 42 58.512 & $-$19 56 58.41 &  0.44^{+0.04}_{-0.04} &  0.18^{+0.06}_{-0.06} & \citet{hel12} & \citet{mor13b}\\
WASP-11 & TYC 2340-1714-1 & 03 09 28.5432 & +30 40 24.853 &  0.46^{+0.02}_{-0.02} &  0.01^{+0.05}_{-0.05} & \citet{wes09b} & \citet{mor13b}\\
WASP-17 & TYC 6787-1927-1 & 15 59 50.947 & $-$28 03 42.33 &  0.46^{+0.03}_{-0.03} & -0.12^{+0.05}_{-0.05} & \citet{and10} & \citet{mor13b}\\
WASP-31 & WASP-31 & 11 17 45.357 & $-$19 03 17.21 &  0.46^{+0.03}_{-0.03} & -0.08^{+0.05}_{-0.05} & \citet{and11b} & \citet{mor13b}\\
WASP-52 & WASP-52 & 23 13 58.76 & +08 45 40.6 &  0.48^{+0.02}_{-0.02} &  0.15^{+0.05}_{-0.05} & \citet{heb13} & \citet{and17}\\
WASP-6 & TYC 6972-75-1 & 23 12 37.7380 & $-$22 40 26.261 &  0.48^{+0.03}_{-0.03} & -0.14^{+0.03}_{-0.03} & \citet{gil09} & \citet{mor13b}\\
HAT-P-46 & GSC 05100-00045 & 18 01 46.608 & $-$02 58 15.43 &  0.49^{+0.08}_{-0.05} &  0.16^{+0.09}_{-0.09} & \citet{har14} & \citet{and17}\\
TrES-4 & NAME TrES-4b & 17 53 13.058 & +37 12 42.36 &  0.50^{+0.04}_{-0.04} &  0.34^{+0.10}_{-0.10} & \citet{man07} & \citet{amm09}\\
CoRoT-5 & CoRoT-5 & 06 45 06.541 & +00 48 54.86 &  0.51^{+0.05}_{-0.03} &  0.04^{+0.05}_{-0.05} & \citet{rau09} & \citet{mor13b}\\
WASP-13 & TYC 2496-1114-1 & 09 20 24.7098 & +33 52 56.717 &  0.51^{+0.05}_{-0.05} &  0.08^{+0.02}_{-0.02} & \citet{ski09} & \citet{gom13}\\
WASP-42 & 2MASS J12515557-4204249 & 12 51 55.571 & $-$42 04 24.99 &  0.51^{+0.03}_{-0.03} &  0.29^{+0.05}_{-0.05} & \citet{len12} & \citet{mor13b}\\
HAT-P-1 & BD+37 4734B & 22 57 46.844 & +38 40 30.33 &  0.52^{+0.02}_{-0.02} &  0.21^{+0.03}_{-0.03} & \citet{bak07b} & \citet{amm09}\\
HAT-P-17 & TYC 2717-417-1 & 21 38 08.7305 & +30 29 19.453 &  0.53^{+0.02}_{-0.02} &  0.05^{+0.03}_{-0.03} & \citet{how12} & \citet{mor13b}\\
OGLE-TR-111 & V* V759 Car & 10 53 17.91 & $-$61 24 20.3 &  0.55^{+0.01}_{-0.01} &  0.22^{+0.15}_{-0.15} & \citet{pon04} & \citet{mor13b}\\
WASP-15 & TYC 7283-1162-1 & 13 55 42.705 & $-$32 09 34.66 &  0.55^{+0.05}_{-0.05} &  0.09^{+0.04}_{-0.04} & \citet{wes09a} & \citet{mor13b}\\
WASP-62 & CPD-64 484 & 05 48 33.5918 & $-$63 59 18.372 &  0.55^{+0.04}_{-0.04} &  0.24^{+0.05}_{-0.05} & \citet{hel12} & \citet{mor13b}\\
HAT-P-39 & GSC 01364-01424 & 07 35 01.979 & +17 49 48.30 &  0.56^{+0.09}_{-0.09} & -0.21^{+0.12}_{-0.12} & \citet{har12} & \citet{and17}\\
WASP-25 & TYC 6706-861-1 & 13 01 26.374 & $-$27 31 19.94 &  0.56^{+0.04}_{-0.04} &  0.06^{+0.03}_{-0.03} & \citet{eno11b} & \citet{mor13b}\\
WASP-22 & TYC 6446-326-1 & 03 31 16.3274 & $-$23 49 10.852 &  0.58^{+0.03}_{-0.02} &  0.26^{+0.03}_{-0.03} & \citet{max10b} & \citet{mor13b}\\
WASP-34 & CD-23 9677 & 11 01 35.8969 & $-$23 51 38.409 &  0.59^{+0.01}_{-0.01} &  0.08^{+0.02}_{-0.02} & \citet{sma11} & \citet{mor13b}\\
WASP-55 & TYC 6125-113-1 & 13 35 01.9530 & $-$17 30 12.523 &  0.59^{+0.04}_{-0.04} &  0.09^{+0.04}_{-0.04} & \citet{hel12} & \citet{mor13b}\\
HAT-P-27 & HAT-P-27 & 14 51 04.189 & +05 56 50.53 &  0.60^{+0.03}_{-0.03} &  0.30^{+0.03}_{-0.03} & \citet{bek11} & \citet{mor13b}\\
WASP-56 & Cl* Melotte 111 AV 561 & 12 13 27.8904 & +23 03 20.459 &  0.60^{+0.04}_{-0.04} &  0.43^{+0.04}_{-0.04} & \citet{fae13} & \citet{mor13b}\\
XO-2 N & TYC 3413-5-1 & 07 48 06.468 & +50 13 32.96 &  0.61^{+0.02}_{-0.02} &  0.37^{+0.07}_{-0.07} & \citet{bur07} & \citet{dam15}\\
OGLE-TR-10 & V* V5125 Sgr & 17 51 28.25 & $-$29 52 34.9 &  0.62^{+0.14}_{-0.14} &  0.28^{+0.10}_{-0.10} & \citet{kon05} & \citet{san06}\\
WASP-54 & BD+00 3088 & 13 41 49.0281 & $-$00 07 41.036 &  0.62^{+0.02}_{-0.02} &  0.00^{+0.03}_{-0.03} & \citet{fae13} & \citet{mor13b}\\
K2-30 & UCAC4 562-007074 & 03 29 22.049 & +22 17 57.75 &  0.63^{+0.03}_{-0.03} &  0.11^{+0.04}_{-0.04} & \citet{joh16} & \citet{lil16}\\
HAT-P-24 & TYC 774-1441-1 & 07 15 18.0149 & +14 15 45.475 &  0.65^{+0.03}_{-0.03} & -0.41^{+0.10}_{-0.10} & \citet{kip10} & \citet{and17}\\
HD 209458 & V* V376 Peg & 22 03 10.7728 & +18 53 03.550 &  0.66^{+0.01}_{-0.01} &  0.03^{+0.02}_{-0.02} & \citet{maz00,cha00,hen00} & \citet{sou08}\\
HAT-P-30 & BD+06 1909 & 08 15 47.9792 & +05 50 12.359 &  0.68^{+0.03}_{-0.03} &  0.12^{+0.03}_{-0.03} & \citet{joh11} & \citet{mor13b}\\
WASP-35 & TYC 4762-714-1 & 05 04 19.6327 & $-$06 13 47.376 &  0.69^{+0.06}_{-0.06} & -0.05^{+0.05}_{-0.05} & \citet{eno11a} & \citet{mor13b}\\
CoRoT-4 & CoRoT-4 & 06 48 46.715 & $-$00 40 21.98 &  0.71^{+0.08}_{-0.08} &  0.15^{+0.06}_{-0.06} & \citet{aig08} & \citet{mor13b}\\
HAT-P-4 & BD+36 2593 & 15 19 57.9275 & +36 13 46.785 &  0.71^{+0.03}_{-0.05} &  0.35^{+0.08}_{-0.08} & \citet{kov07} & \citet{amm09}\\
TrES-1 & NAME TrES-1b & 19 04 09.844 & +36 37 57.54 &  0.73^{+0.05}_{-0.04} &  0.06^{+0.05}_{-0.05} & \citet{alo04} & \citet{san06}\\
HAT-P-41 & TYC 488-2442-1 & 19 49 17.4383 & +04 40 20.763 &  0.75^{+0.10}_{-0.10} &  0.13^{+0.05}_{-0.05} & \citet{har12} & \citet{tsa14}\\
CoRoT-9 & CoRoT-9 & 18 43 08.810 & +06 12 14.89 &  0.81^{+0.07}_{-0.07} & -0.02^{+0.03}_{-0.03} & \citet{dee10} & \citet{mor13b}\\
CoRoT-12 & CoRoT-12 & 06 43 03.762 & $-$01 17 47.12 &  0.85^{+0.07}_{-0.06} &  0.17^{+0.14}_{-0.14} & \citet{gil10} & \citet{mor13b}\\
WASP-16 & TYC 6147-229-1 & 14 18 43.922 & $-$20 16 31.85 &  0.86^{+0.04}_{-0.08} &  0.13^{+0.02}_{-0.02} & \citet{list09} & \citet{mor13b}\\
XO-1 & BD+28 2507 & 16 02 11.8470 & +28 10 10.421 &  0.86^{+0.07}_{-0.07} & -0.01^{+0.05}_{-0.05} & \citet{mcc06} & \citet{amm09}\\
WASP-1 & TYC 2265-107-1 & 00 20 40.077 & +31 59 23.79 &  0.87^{+0.06}_{-0.05} &  0.23^{+0.03}_{-0.03} & \citet{col07} & \citet{mor13b}\\
WASP-2 & WASP-2 & 20 30 54.130 & +06 25 46.37 &  0.87^{+0.09}_{-0.09} &  0.02^{+0.05}_{-0.05} & \citet{col07} & \citet{mor13b}\\
WASP-76 & BD+01 316 & 01 46 31.8590 & +02 42 02.065 &  0.88^{+0.03}_{-0.03} &  0.36^{+0.04}_{-0.04} & \citet{wes16} & \citet{and17}\\
WASP-23 & GSC 07635-01376 & 06 44 30.65 & $-$42 45 41.0 &  0.91^{+0.09}_{-0.10} &  0.05^{+0.06}_{-0.06} & \citet{tri11} & \citet{mor13b}\\
WASP-28 & 2MASS J23342787-0134482 & 23 34 27.881 & $-$01 34 48.13 &  0.91^{+0.04}_{-0.04} & -0.12^{+0.03}_{-0.03} & \citet{and15} & \citet{mor13b}\\
WASP-41 & TYC 7247-587-1 & 12 42 28.497 & $-$30 38 23.55 &  0.92^{+0.05}_{-0.05} &  0.06^{+0.02}_{-0.02} & \citet{max11} & \citet{mor13b}\\
WASP-79 & CD-30 1812 & 04 25 29.0162 & $-$30 36 01.603 &  0.92^{+0.08}_{-0.08} &  0.19^{+0.10}_{-0.10} & \citet{sma12} & \citet{mor13b}\\
WASP-78 & TYC 5889-271-1 & 04 15 01.5044 & $-$22 06 59.109 &  0.93^{+0.08}_{-0.08} & -0.07^{+0.05}_{-0.05} & \citet{sma12} & \citet{mor13b}\\
WASP-7 & HD 197286 & 20 44 10.2190 & $-$39 13 30.894 &  0.95^{+0.13}_{-0.13} &  0.12^{+0.09}_{-0.09} & \citet{hel09b} & \citet{mor13b}\\
WASP-44 & GSC 05264-00740 & 00 15 36.770 & $-$11 56 17.30 &  0.96^{+0.07}_{-0.07} &  0.17^{+0.06}_{-0.06} & \citet{and12} & \citet{and17}\\
WASP-58 & TYC 3525-76-1 & 18 18 48.2530 & +45 10 19.257 &  0.98^{+0.08}_{-0.08} & -0.09^{+0.04}_{-0.04} & \citet{heb13} & \citet{and17}\\
HAT-P-35 & TYC 203-1079-1 & 08 13 00.1827 & +04 47 13.382 &  1.01^{+0.03}_{-0.03} &  0.12^{+0.03}_{-0.03} & \citet{bak12} & \citet{mor13b}\\
OGLE-TR-182 & OGLE-TR 182 & 11 09 18.71 & $-$61 05 42.9 &  1.01^{+0.15}_{-0.15} &  0.37^{+0.08}_{-0.08} & \citet{pon08} & \citet{pon08}\\
OGLE-TR-211 & 2MASS J10401438-6227201 & 10 40 14.39 & $-$62 27 20.2 &  1.02^{+0.20}_{-0.20} &  0.11^{+0.10}_{-0.10} & \citet{uda08} & \citet{uda08}\\
WASP-26 & TYC 5839-876-1 & 00 18 24.7001 & $-$15 16 02.287 &  1.02^{+0.02}_{-0.02} &  0.16^{+0.02}_{-0.02} & \citet{sma10} & \citet{mor13b}\\
WASP-24 & TYC 339-329-1 & 15 08 51.7355 & +02 20 35.953 &  1.03^{+0.03}_{-0.04} &  0.09^{+0.04}_{-0.04} & \citet{str10} & \citet{mor13b}\\
HAT-P-6 & TYC 3239-992-1 & 23 39 05.8108 & +42 27 57.502 &  1.04^{+0.12}_{-0.12} & -0.08^{+0.11}_{-0.11} & \citet{noy08} & \citet{amm09}\\
WASP-45 & TYC 6996-583-1 & 00 20 56.9941 & $-$35 59 53.756 &  1.04^{+0.06}_{-0.06} &  0.43^{+0.06}_{-0.06} & \citet{and12} & \citet{mor13b}\\
HAT-P-42 & GSC 00232-01451 & 09 01 22.648 & +06 05 49.99 &  1.07^{+0.09}_{-0.09} &  0.34^{+0.05}_{-0.05} & \citet{boi13} & \citet{and17}\\
WASP-82 & TYC 88-57-1 & 04 50 38.5600 & +01 53 38.088 &  1.11^{+0.04}_{-0.04} &  0.18^{+0.04}_{-0.04} & \citet{wes16} & \citet{and17}\\
WASP-19 & GSC 08181-01711 & 09 53 40.077 & $-$45 39 33.06 &  1.13^{+0.04}_{-0.04} &  0.26^{+0.05}_{-0.05} & \citet{heb10} & \citet{mor13b}\\
WASP-75 & GSC 05816-01135 & 22 49 32.568 & $-$10 40 31.93 &  1.14^{+0.05}_{-0.05} &  0.24^{+0.03}_{-0.03} & \citet{gom13} & \citet{and17}\\
CoRoT-1 & CoRoT-1 & 06 48 19.172 & $-$03 06 07.68 &  1.16^{+0.13}_{-0.13} &  0.03^{+0.04}_{-0.04} & \citet{bar08} & \citet{mor13b}\\
OGLE-TR-132 & V* V742 Car & 10 50 34.72 & $-$61 57 25.9 &  1.16^{+0.14}_{-0.13} &  0.37^{+0.07}_{-0.07} & \citet{bou04} & \citet{gil07}\\
WASP-95 & CD-48 14223 & 22 29 49.7348 & $-$48 00 11.012 &  1.16^{+0.10}_{-0.04} &  0.22^{+0.03}_{-0.03} & \citet{hel14} & \citet{and17}\\
WASP-47 & EPIC 206103150 & 22 04 48.731 & $-$12 01 07.99 &  1.18^{+0.58}_{-0.38} &  0.36^{+0.05}_{-0.05} & \citet{hel12} & \citet{mor13b}\\
HD 189733 & HD 189733 & 20 00 43.7128 & +22 42 39.074 &  1.19^{+0.06}_{-0.06} &  0.03^{+0.08}_{-0.08} & \citet{bou05} & \citet{mor13b}\\
WASP-4 & 2MASS J23341508-4203411 & 23 34 15.082 & $-$42 03 41.14 &  1.22^{+0.01}_{-0.01} &  0.03^{+0.03}_{-0.03} & \citet{wil08} & \citet{mor13b}\\
TrES-2 & Kepler-1b & 19 07 14.035 & +49 18 59.07 &  1.25^{+0.07}_{-0.07} &  0.06^{+0.08}_{-0.08} & \citet{odo06} & \citet{amm09}\\
OGLE-TR-113 & V* V752 Car & 10 52 24.40 & $-$61 26 48.5 &  1.26^{+0.16}_{-0.16} &  0.03^{+0.06}_{-0.06} & \citet{bou04} & \citet{mor13b}\\
WASP-97 & CD-56 324 & 01 38 25.0569 & $-$55 46 19.511 &  1.33^{+0.05}_{-0.05} &  0.31^{+0.04}_{-0.04} & \citet{hel14} & \citet{and17}\\
WASP-12 & WASP-12 & 06 30 32.7943 & +29 40 20.287 &  1.35^{+0.07}_{-0.06} &  0.21^{+0.04}_{-0.04} & \citet{heb09} & \citet{mor13b}\\
HAT-P-8 & HAT-P-8 & 22 52 09.8629 & +35 26 49.605 &  1.44^{+0.17}_{-0.15} &  0.07^{+0.04}_{-0.04} & \citet{lat09} & \citet{mor13b}\\
OGLE-TR-056 & V* V5157 Sgr & 17 56 35.51 & $-$29 32 21.2 &  1.44^{+0.19}_{-0.18} &  0.25^{+0.08}_{-0.08} & \citet{kon03} & \citet{san06}\\
WASP-72 & CD-30 1019 & 02 44 09.6110 & $-$30 10 08.570 &  1.45^{+0.06}_{-0.05} &  0.15^{+0.06}_{-0.06} & \citet{gil13} & \citet{and17}\\
WASP-50 & TYC 5290-462-1 & 02 54 45.1351 & $-$10 53 53.037 &  1.51^{+0.09}_{-0.09} &  0.13^{+0.03}_{-0.03} & \citet{gil11} & \citet{mor13b}\\
KOI-1257 & Kepler-420 & 19 24 54.043 & +44 55 38.58 &  1.55^{+0.37}_{-0.37} &  0.22^{+0.04}_{-0.04} & \citet{san14} & \citet{san14}\\
WASP-5 & GSC 08018-00199 & 23 57 23.759 & $-$41 16 37.74 &  1.62^{+0.13}_{-0.10} &  0.17^{+0.06}_{-0.06} & \citet{and08} & \citet{mor13b}\\
WASP-77 A & BD-07 436A & 02 28 37.2266 & $-$07 03 38.366 &  1.71^{+0.06}_{-0.06} &  0.07^{+0.03}_{-0.03} & \citet{max13} & \citet{mor13b}\\
K2-34 & TYC 1391-121-1 & 08 30 18.9080 & +22 14 09.313 &  1.79^{+0.13}_{-0.13} &  0.11^{+0.04}_{-0.04} & \citet{hir16} & \citet{lil16}\\
HAT-P-7 & BD+47 2846 & 19 28 59.3534 & +47 58 10.229 &  1.81^{+0.02}_{-0.02} &  0.31^{+0.07}_{-0.07} & \citet{pal08} & \citet{amm09}\\
TrES-3 & NAME TrES-3b & 17 52 07.020 & +37 32 46.18 &  1.84^{+0.07}_{-0.08} & -0.10^{+0.19}_{-0.19} & \citet{odo07} & \citet{amm09}\\
WASP-100 & CPD-64 356 & 04 35 50.3297 & $-$64 01 37.316 &  1.87^{+0.11}_{-0.11} & -0.30^{+0.12}_{-0.12} & \citet{hel14} & \citet{and17}\\
WASP-73 & HD 202678 & 21 19 47.9070 & $-$58 08 55.951 &  1.87^{+0.07}_{-0.06} &  0.20^{+0.02}_{-0.02} & \citet{del14} & \citet{and17}\\
WASP-3 & TYC 2636-195-1 & 18 34 31.6241 & +35 39 41.488 &  1.92^{+0.06}_{-0.06} & -0.02^{+0.08}_{-0.08} & \citet{pol08} & \citet{mon12}\\
WASP-37 & GSC 00326-00658 & 14 47 46.560 & +01 03 53.85 &  1.93^{+0.18}_{-0.18} & -0.23^{+0.05}_{-0.05} & \citet{sim11} & \citet{and17}\\
HATS-1 & HATS-1 & 11 42 06.080 & $-$23 21 17.44 &  1.94^{+0.27}_{-0.21} & -0.04^{+0.04}_{-0.04} & \citet{pen13} & \citet{and17}\\
WASP-61 & TYC 6469-1972-1 & 05 01 11.9193 & $-$26 03 14.973 &  2.00^{+0.17}_{-0.17} & -0.38^{+0.11}_{-0.11} & \citet{hel12} & \citet{and17}\\
HAT-P-23 & TYC 1632-1396-1 & 20 24 29.7254 & +16 45 43.812 &  2.11^{+0.12}_{-0.12} &  0.16^{+0.03}_{-0.03} & \citet{bak11} & \citet{tsa14}\\
WASP-71 & TYC 30-116-1 & 01 57 03.2070 & +00 45 31.865 &  2.11^{+0.08}_{-0.08} &  0.37^{+0.04}_{-0.04} & \citet{smi13} & \citet{mor13b}\\
HAT-P-14 & TYC 3086-152-1 & 17 20 27.8775 & +38 14 31.911 &  2.19^{+0.06}_{-0.06} &  0.17^{+0.07}_{-0.07} & \citet{tor10} & \citet{sou15}\\
WASP-36 & GSC 05442-00759 & 08 46 19.298 & $-$08 01 37.01 &  2.24^{+0.07}_{-0.07} & -0.01^{+0.05}_{-0.05} & \citet{smi12} & \citet{mor13b}\\
Kepler-17 & KOI-203 & 19 53 34.866 & +47 48 54.02 &  2.26^{+0.10}_{-0.10} &  0.26^{+0.10}_{-0.10} & \citet{des11} & \citet{bon12}\\
WASP-8 & TYC 7522-505-1 & 23 59 36.0711 & $-$35 01 52.920 &  2.26^{+0.08}_{-0.09} &  0.29^{+0.03}_{-0.03} & \citet{que10} & \citet{mor13b}\\
HAT-P-22 & HD 233731 & 10 22 43.5927 & +50 07 42.060 &  2.37^{+0.07}_{-0.07} &  0.28^{+0.05}_{-0.05} & \citet{bak11} & \citet{sou15}\\
WASP-66 & TYC 7193-1804-1 & 10 32 53.9926 & $-$34 59 23.452 &  2.43^{+0.14}_{-0.14} &  0.05^{+0.05}_{-0.05} & \citet{hel12} & \citet{mor13b}\\
WASP-99 & CD-50 777 & 02 39 35.4435 & $-$50 00 28.875 &  2.46^{+0.11}_{-0.11} &  0.27^{+0.06}_{-0.06} & \citet{hel14} & \citet{and17}\\
CoRoT-10 & CoRoT-10 & 19 24 15.296 & +00 44 45.99 &  2.56^{+0.15}_{-0.15} &  0.06^{+0.09}_{-0.09} & \citet{bon10} & \citet{mor13b}\\
WASP-38 & HD 146389 & 16 15 50.3646 & +10 01 57.280 &  2.60^{+0.06}_{-0.06} &  0.06^{+0.04}_{-0.04} & \citet{bar11} & \citet{mor13b}\\
CoRoT-11 & CoRoT-11 & 18 42 44.947 & +05 56 15.70 &  2.67^{+0.39}_{-0.39} &  0.04^{+0.03}_{-0.03} & \citet{gan10} & \citet{tsa14}\\
Qatar-2 & Qatar 2 & 13 50 37.409 & $-$06 48 14.41 &  2.80^{+0.06}_{-0.06} &  0.09^{+0.17}_{-0.17} & \citet{bry12} & \citet{and17}\\
Kepler-43 & Kepler-43 & 19 00 57.810 & +46 40 05.62 &  3.11^{+0.25}_{-0.25} &  0.33^{+0.11}_{-0.11} & \citet{bon12} & \citet{bon12}\\
HD 17156 & HD 17156 & 02 49 44.4873 & +71 45 11.630 &  3.13^{+0.08}_{-0.08} &  0.23^{+0.04}_{-0.04} & \citet{fis07} & \citet{amm09}\\
WASP-10 & GSC 02752-00114 & 23 15 58.299 & +31 27 46.28 &  3.18^{+0.13}_{-0.11} &  0.04^{+0.05}_{-0.05} & \citet{chr09} & \citet{mor13b}\\
HAT-P-34 & HD 351766 & 20 12 46.8851 & +18 06 17.431 &  3.28^{+0.21}_{-0.21} &  0.08^{+0.05}_{-0.05} & \citet{bak12} & \citet{tsa14}\\
CoRoT-2 & CoRoT-2 & 19 27 06.496 & +01 23 01.38 &  3.30^{+0.21}_{-0.21} & -0.09^{+0.07}_{-0.07} & \citet{alo08} & \citet{mor13b}\\
WASP-32 & TYC 2-1155-1 & 00 15 50.8103 & +01 12 01.592 &  3.88^{+0.08}_{-0.08} &  0.28^{+0.10}_{-0.10} & \citet{max10a} & \citet{mor13b}\\
HD 80606 & HD 80606 & 09 22 37.5769 & +50 36 13.430 &  4.02^{+0.11}_{-0.11} &  0.32^{+0.09}_{-0.09} & \citet{nae01} & \citet{san04}\\
HAT-P-20 & HAT-P-20 & 07 27 39.950 & +24 20 11.49 &  7.31^{+0.19}_{-0.19} &  0.12^{+0.15}_{-0.15} & \citet{bak11} & \citet{mor13b}\\
HAT-P-2 & HD 147506 & 16 20 36.3576 & +41 02 53.107 &  9.00^{+0.24}_{-0.24} &  0.04^{+0.05}_{-0.05} & \citet{bak07a} & \citet{tsa14}\\
WASP-18 & HD 10069 & 01 37 25.0332 & $-$45 40 40.373 & 10.16^{+0.37}_{-0.37} &  0.19^{+0.05}_{-0.05} & \citet{hel09a} & \citet{mor13b}\\
XO-3 & TYC 3727-1064-1 & 04 21 52.7053 & +57 49 01.868 & 13.06^{+0.65}_{-0.65} & -0.08^{+0.04}_{-0.04} & \citet{joh08} & \citet{tsa14}\\
CoRoT-3 & CoRoT-3 & 19 28 13.265 & +00 07 18.62 & 22.08^{+1.02}_{-1.02} &  0.14^{+0.04}_{-0.04} & \citet{del08} & \citet{tsa14}\\
\enddata
\end{deluxetable*}
\end{longrotatetable}

\begin{rotatetable*}
\begin{deluxetable*}{ccllDDcc}
\tabletypesize{\tiny}
\tablecaption{Brown dwarfs and Low-mass Stars that have Doppler-inferred Masses and Transit Solar-type Stars\label{tbl-2}}
\tablehead{
\colhead{System} & \colhead{Simbad} & \colhead{R.A.} & \colhead{Decl.} & \twocolhead{Mass} & \twocolhead{Metallicity} & \colhead{Discovery} & \colhead{Parameter} \\
\colhead{} & \colhead{Name} & \colhead{} & \colhead{} & \twocolhead{} & \twocolhead{} & \colhead{Reference} & \colhead{Reference\tablenotemark{a}} \\
\colhead{} & \colhead{} & \colhead{(h m s)} & \colhead{(d m s)} & \twocolhead{($M_{\mathrm{Jup}}$)} & \twocolhead{} & \colhead{} & \colhead{}
}
\decimals
\startdata
Kepler-39 & KOI-423 & 19 47 50.468 & +46 02 03.43 & $20.1^{+1.2}_{-1.3}$ & $0.1^{+0.14}_{-0.14}$ & \citet{bou11a} & \citet{bon15}\\
KELT-1 & TYC 2785-2130-1 & 00 01 26.9206 & +39 23 01.773 & $27.37^{+0.92}_{-0.92}$ & $0.052^{+0.079}_{-0.079}$ & \citet{siv12} & \\
EPIC 219388192 & UCAC4 366-166973 & 19 17 34.026 & $-$16 52 17.75 & $36.5^{+0.09}_{-0.09}$ & $0.03^{+0.08}_{-0.08}$ & \citet{cur16} & \citet{now17}\\
KOI-205 & KOI-205 & 19 41 59.197 & +42 32 16.40 & $40.8^{+1.5}_{-1.1}$ & $0.18^{+0.12}_{-0.12}$ & \citet{dia13} & \citet{bon15}\\
CoRoT-33 & 2MASS J18383391+0537287 & 18 38 33.917 & +05 37 28.77 & $59.^{+1.7}_{-1.8}$ & $0.44^{+0.1}_{-0.1}$ & \citet{csi15} & \\
KOI-415 & KOI-415 & 19 33 13.452 & +41 36 22.94 & $62.14^{+2.69}_{-2.69}$ & $-0.24^{+0.11}_{-0.11}$ & \citet{mou13} & \\
WASP-30 & WASP-30 & 23 53 38.0552 & $-$10 07 05.107 & $62.5^{+1.2}_{-1.2}$ & $0.083^{+0.05}_{-0.069}$ & \citet{and11a} & \citet{tri13} \\
CoRoT-15 & CoRoT-15 & 06 28 27.819 & +06 11 10.54 & $63.3^{+4.1}_{-4.1}$ & $0.1^{+0.2}_{-0.2}$ & \citet{bou11b} & \\
EPIC 201702477 & UCAC2 33047398 & 11 40 57.792 & +03 40 53.70 & $66.9^{+1.7}_{-1.7}$ & $-0.164^{+0.053}_{-0.053}$ & \citet{bay17} & \\
KOI-189 & KOI-189 & 18 59 31.191 & +49 16 01.18 & $78.^{+3.4}_{-3.4}$ & $-0.115^{+0.099}_{-0.099}$ & \citet{dia14} & \\
EBLM J0555-57A & CD-57 1311 & 05 55 32.6868 & $-$57 17 26.064 & $85.2^{+3.9}_{-4}$ & $-0.24^{+0.16}_{-0.16}$ & \citet{von17} & \\
TYC 7760-484-1 & CD-39 7570 & 12 19 21.0385 & $-$39 51 25.575 & $95.4^{+2.5}_{-1.9}$ & $-0.209^{+0.075}_{-0.07}$ & \citet{tri13} & \\
OGLE-TR-122 & V* V817 Car & 11 06 51.89 & $-$60 51 45.9 & $96.^{+9}_{-9}$ & $0.15^{+0.36}_{-0.36}$ & \citet{pon05} & \\
1SWASPJ234318.41+295556.5 & BD+29 4980 & 23 43 18.4125 & +29 55 56.696 & $100.^{+7}_{-7}$ & $0.07^{+0.17}_{-0.008}$ & \citet{cha16} & \\
CoRoT 101186644 & 2MASS J19265907+0029061 & 19 26 59.078 & +00 29 06.19 & $100.^{+12}_{-12}$ & $0.2^{+0.2}_{-0.2}$ & \citet{tal13} & \\
KOI-686 & KOI-686 & 19 47 21.783 & +43 38 49.64 & $103.4^{+4.8}_{-4.8}$ & $-0.06^{+0.13}_{-0.13}$ & \citet{dia14} & \\
HATS550-016 & GSC 06465-00602 & 04 48 23.318 & $-$24 50 16.88 & $115.^{+6}_{-5}$ & $-0.6^{+0.06}_{-0.06}$ & \citet{zho14} & \\
HAT-TR-205-013 & NAME HAT-TR-205-013 & 23 08 08.3420 & +33 38 03.963 & $130.^{+10}_{-10}$ & $-0.2^{+0.2}_{-0.2}$ & \citet{bea07} & \\
HATS551-021 & GSC 06493-00315 & 05 42 49.120 & $-$25 59 47.49 & $138.^{+5}_{-15}$ & $-0.4^{+0.1}_{-0.1}$ & \citet{zho14} & \\
KIC 1571511 & KOI-362 & 19 23 59.256 & +37 11 57.19 & $150.^{+4}_{-5}$ & $0.37^{+0.08}_{-0.08}$ & \citet{ofi12} & \\
HATS551-019 & TYC 6493-290-1 & 05 40 46.1695 & $-$24 55 35.189 & $180.^{+10}_{-10}$ & $-0.4^{+0.1}_{-0.1}$ & \citet{zho14} & \\
TYC 3700-1739-1 & TYC 3700-1739-1 & 02 40 51.519 & +52 45 06.27 & $197.^{+15}_{-15}$ & $-0.05^{+0.17}_{-0.17}$ & \citet{eig16} & \\
T-Lyr1-01662 & TYC 3545-371-1 & 18 59 02.8586 & +48 36 35.557 & $207.^{+13}_{-13}$ & $-0.5^{+0.2}_{-0.2}$ & \citet{fer09} & \\
HATS553-001 & GSC 05946-00892 & 06 16 00.656 & $-$21 15 23.82 & $210.^{+20}_{-10}$ & $-0.1^{+0.2}_{-0.2}$ & \citet{zho14} & \\
Kepler-16 & Kepler-16 & 19 16 18.175 & +51 45 26.76 & $212.17^{+0.68}_{-0.69}$ & $-0.3^{+0.2}_{-0.2}$ & \citet{doy11} & \\
T-Lyr0-08070 & TYC 3121-1659-1 & 19 19 03.7120 & +38 40 56.775 & $251.^{+20}_{-20}$ & $-0.5^{+0.2}_{-0.2}$ & \citet{fer09} & \\
HD 213572 & HD 213572 & 22 32 00.1305 & +27 14 09.767 & $300.^{+10}_{-13}$ & $-0.065^{+0.048}_{-0.05}$ & \citet{cha14} & \\
\enddata
\tablenotetext{a}{Only included if it is different than the discovery reference.}
\end{deluxetable*}
\end{rotatetable*}

\listofchanges
\end{document}